\documentstyle[aps,prl,multicol]{revtex}

\input epsf

\def\date#1{\author{\small(#1)}}
\def\abstract#1{\author{\parbox[t]{5.5in}{\small#1}}\par\maketitle}

\def\f.#1.{{\bf #1}}
\def\mb.#1.{\bbox{#1}}

\def\opn{\begin{equation}} \def\cls{\end{equation}}
\def\opa{\begin{eqnarray}} \def\cla{\end{eqnarray}}
\def\opbib{\begin{references}} \def\clbib{\end{references}}
\def\bb#1{\bibitem{#1}}
\def\qno;#1;{\label{#1}\end{equation}} \def\qna;#1;{\label{#1}\end{eqnarray}}
\def\rf;#1;{(\ref{#1})}
\def\dels#1 {\nabla\kern -1.5pt_{#1}\kern 1.5pt}
 
\def\avv#1{\langle #1\rangle}
\def\sgn{{\rm sign}} 

\def\suspend{\end{multicols}\vspace*{-0.5cm} \noindent \rule{8.65cm}{.02cm}}
\def\resume{\hskip 9.3cm \rule{8.65cm}{.02cm} \begin{multicols}{2}\vspace*{-0.8cm} \noindent}
\def\umo{\rlap{\"{$\,$\ }} \,\,\,\kern -.3 cm o}

\newcount\hours \newcount\minutes \newcount\a \newcount\b
\def\gmt{\hours = \time \divide\hours by 60 \a =\hours \multiply \a by 60
\minutes = \time \advance \minutes by -\a
{\ifnum\hours<10 0\fi}\number\hours
{\ifnum\minutes<10 0\fi}\number\minutes}
\def\gday{\b=\year \advance \b by -1900
\number\b{\ifnum\month<10 0\fi}\number\month{\ifnum\day<10 0\fi}\number\day}
\def\today{\number\day\ 
\ifcase\month\or January\or February\or March\or April\or May\or 
June\or July\or August\or September\or October\or November\or December\fi
\ \number\year}

\def\al{\alpha} \def\be{\beta} \def\ga{\gamma} \def\de{\delta}
\def\ep{\epsilon}  \def\et{\eta}  
 \def\la{\lambda}  \def\si{\sigma} \def\ta{\tau}
  \def\ch{\chi} 
 \def\De{\Delta}

\def\casefr#1/#2 {\case{#1}{#2}}
\def\part{\partial}

  \def\prop{\propto}

 \def\br{\f.r.}  
\def\bk{\f.k.}

\def\sump#1{\lower .15in\hbox{$\stackrel{\displaystyle{\sum}'} {\scriptstyle #1}$}}

\def\tq{\tilde q} \def\hu{\hat u} 
\def\resume{\hbox to 18cm {\hfill \rule{8.65cm}{.02cm}}\begin{multicols}{2}\vspace*{-0.8cm} \noindent}

\begin{document}

\title{\vbox to 0pt {\vskip -1cm \rlap{\hbox to \textwidth {\rm{\small For Physica D, Proceedings of CNLS Workshop on Anomalous Distributions, etc, Santa Fe NM, Nov 6-9, 2002\hfill}}}}Turbulence and Tsallis Statistics}

\author{Toshiyuki Gotoh$^{1}$ and Robert H. Kraichnan$^{2}$}

\address{$^1$Department of Systems Engineering, Nagoya Institute of Technology, Showa-ku, Nagoya 466, Japan}
\address{$^2$369 Montezuma \#108, Santa Fe, NM 87501-2626}

\date{20 May 2003}

\abstract{Fully-developed incompressible Navier-Stokes turbulence in three dimensions is a dissipative dynamical system that exhibits strong departure from absolute equilibrium. Nevertheless, several kinds of representation by Tsallis equilibria have been proposed. The symmetry of the contributions to the kinetic energy from the degrees of freedom of the flow must be broken in order to construct turbulence applications. Tsallis representations of turbulence involve the extrapolation of results for systems with a single degree of freedom, a procedure that requires critical examination. Theories of Tsallis statistics for acceleration of fluid particles are compared with computer simulations of the pressure gradient field in the present paper. Applications of Tsallis formulations to statistics of longitudinal velocity differences are analyzed. No flavor of equilibrium statistics based on the kinetic energy can encompass the dynamically fundamental turbulent energy cascade, which requires correlation of triads of degrees of freedom at different scales.
}
\vskip 4mm

{\hskip 16mm PACS: 47.27.Gs; 05.45.-a; 47.53.+n; 05.20.-y}

{\hskip 16mm Keywords: turbulence; Tsallis statistics; turbulent pressure; turbulent acceleration}

\vskip 4mm

\begin{multicols}{2}

\section*{1. Introduction}

The search for something that is maximized or minimized by turbulent flows extends back over more than a half-century. Non-extensive entropy has joined a varied set of candidates.

Fully developed incompressible Navier-Stokes (NS) turbulence is a dissipative dynamical system with very many degrees of freedom in a state of strong departure from absolute statistical equilibrium. The disequilibrium is expressed by several features. In isotropic NS turbulence in three dimensions (3D), the wavenumber energy spectrum $E(k)$ at high Reynolds numbers goes approximately like $k^{-5/3}$ in the inertial range of $k$, which extends between the low $k$ where any driving occurs and the high $k$ where viscous dissipation is strong. In contrast, Boltzmann-Gibbs equilibrium based on kinetic-energy yields $E(k) \prop k^2$. Moreover, there is strong energy-transferring coupling within the inertial range between $k$ that differ by a factor greater than 2. If one thinks of a temperature local in $k$, this means strong coupling among degrees of freedom whose local temperatures differ by a factor like 2.

In 1988 Tsallis proposed a generalization of statistical mechanics that can describe non-extensive statistical equilibrium, in contrast to other generalizations, like that of R\'{e}nyi, that preserve the additivity of the entropies of independent subsystems \cite{1}. The Boltzmann-Gibbs equilibrium is recovered as a limiting case. Tsallis statistics have been formulated for a variety of systems including fluid turbulence. The theoretical formulations for 3D NS turbulence assume isotropy. The results have been compared not only with approximately isotropic direct numerical simulation (DNS) but also with experiments that have more complex geometries. The key analysis typically is performed for systems very much simpler than turbulence: one-dimensional maps and dynamical systems consisting of a single degree of freedom that obeys a modified Langevin equation.

\section*{2. Tsallis statistics}

A number of generalized definitions of entropy have been proposed that reduce to Boltzmann-Gibbs entropy in a limit. The entropy proposal by R\'{e}nyi \cite{1} was motivated by information-theoretic considerations. It yields additivity of the entropies of independent subsystems. Tsallis statistics are based upon the entropy formula
\opn
S_q = {1\over(q-1)}\left( 1- \sum_i p_i^q \right), \quad \sum_i p_i = 1,
\qno;1;
where $q$ is a parameter of the theory and the sum is over the probabilities $p_i$ of states $i$ that assign values to a complete set of real dynamic variables $y_n$.

If $q \ne 1$, $S_q$ is not the sum of contributions from each $i$. Independent subsystems give interlocked contributions to $S_q$, yielding non-extensive equilibrium. The limit $q \to 1$ yields the Boltzmann formula $S_q = - \sum p_i\ln p_i$. 

Let $E_i$ be the energy of state $i$. The extreme value of $S_q$ under the generalized energy constraint
\opn
\sum_i p_i^q E_i = U_q\sum_i p_i^q,
\qno;2;
is realized by the probabilities
\opn
p_i = {1\over Z_q} [1 + (q-1)\be E_i]^{-1/(q-1)}.
\qno;3;
Here $k_B\be$ is an inverse temperature parameter whose value is determined by the value taken for $U_q$, $k_B$ is Boltzmann's constant and
\opn
Z_q = \sum_i [1 + (q-1)\be E_i]^{-1/(q-1)}.
\qno;4;

The Lagrange multiplier for the energy constraint is written in the form $(q-1)\be$ so that the limit $q \to 1$ yields the Boltzmann-Gibbs values
\opn
p_i = \exp(-\be E_i)/Z_B, \quad Z_B = \sum_i\exp(-\be E_i).
\qno;5;

Consider the case in which (a) each $E_i$ is a sum of energy contributions $e_n$ from the $y_n$; (b) the dynamics conserve energy; (c) there is a Liouville property $\sum_n \part\dot y_n/\part y_n = 0$. In this case it is self-consistent to take uniform density of states in the phase space of the $y_n$. The Boltzmann-Gibbs formula \rf;5; then yields an explicit result for the probability density function (PDF) of $y_n$:
\opn
P_n(y_n) = \exp(-\be e_n)/S_n, \quad S_n = \int_{-\infty}^\infty \exp(-\be e_n)dy_n.
\qno;6;
If $q \ne 1$, $E_i$ that are sums of contributions from each $y_n$ do not yield separation of the PDFs of distinct $y_n$. In most cases $P_n(y_n)$ can be obtained only by integration over all the $y_{m \ne n}$.

The particular case
\opn
E_i = [u^2/2]_i \equiv \left[\sum_n y_n^2/2\right]_i \quad n = 1,2,...N,
\qno;7;
is appropriate to a suitably constructed truncated representation of Euler or NS flow. In this case only, the constant energy surfaces are hyperspherical surfaces in the phase space of the $y_n$. The Boltzmann formula \rf;6; immediately yields Gaussian $P_n(y_n)$.

If $q \ne 1$, \rf;3; and \rf;7; yield complete symmetry in $n$. An explicit result for $P_n$ can be extracted if, in addition, $N \to \infty$. Then the distribution of each $y_n$ on the energy shell tends to the Gaussian
\opn
[P_n(y_n)]_i = {1\over u}\sqrt{N\over 2\pi}\exp(-{\textstyle\frac{1}{2}} N y_n^2/u^2).
\qno;8;
With the use of \rf;3;, \rf;8; can be integrated over $u$ to give
\[ P_n(y_n) = {1\over Z_q}{\sqrt{N\over 2\pi}} \nonumber \]
\opn
\times \int_0^\infty [1 + {\textstyle\frac{1}{2}} \be(q-1)u^2]^{-1/(q-1)} \exp(-{\textstyle\frac{1}{2}} N y_n^2/u^2) {du\over u}.
\qno;9;

The canonical distribution \rf;5; is realized if the conservative system is put in contact with a heat bath and then removed. This implies unique stability properties for the canonical distribution that can be demonstrated as follows. Consider two systems $A$ and $B$ that are initially independent with PDFs $P_A(E_A)$ and $P_A(E_B)$ that are functions only of the system energies. The initial PDF for the combined system is then $P_A(E_A)P_B(E_B)$. Now let the systems be coupled by arbitrary couplings that conserve the total energy $E_A + E_B$ and preserve the Liouville property. The condition that the total system PDF be invariant under these couplings is that the PDF keeps the form $P_A(E_A)P_B(E_B)$ and is time invariant. But since there are arbitrary couplings, the PDF must also have the form $P_{A+B}(E_A + E_B)$. Equality of these two forms has the canonical distribution as its unique solution, a fact easily verified by taking the derivative with respect to $E_A$ and separately with respect to $E_B$, then rearranging terms.

Tsallis $q \ne 1$ statistics for a conservative system with Liouville property do not have the robust stability properties of the canonical distribution. A necessary condition for $q \ne 1$ is that the system be isolated from its environment or the environment be specially constructed.

Dissipation destroys the Liouville property. Equilibrium statistics can be extended to some dissipative systems, such as a single degree of freedom that obeys a Langevin equation with suitable relation between damping and random forcing. If particular assignments of effective energy are made, dissipationless equilibrium Tsallis statistics  yields exact PDFs for some dissipative systems. Examples of such systems, with a single degree of freedom, will be discussed in Sec. 4. NS turbulence at high Reynolds numbers is a dissipative system with many strongly interacting degrees of freedom. In the forced steady state, the forced degrees of freedom are separated from those where the dissipation occurs by a multistage energy cascade. Any attempt to describe NS turbulence by equilibrium statistics must be considered courageous.

\section*{3. Application to Navier-Stokes Flow}

An underlying fact is that the quadratic inviscid constants of motion give no hint of the rich dynamics that emanate from the NS equation. For example there is nothing to imply the crucially important Kelvin circulation theorem.

The exploration of the Boltzmann-Gibbs and Tsallis statistics of Sec. 2 for NS flow begins with the choice of a representation basis for the flow. There is an infinity of representation bases, constrained by the boundary conditions on the flow. The DNS to be described in what follows uses cyclic boundary conditions on a cube. This is appropriate for spatially homogeneous flow. The representation most immediately suggested by homogeneity is expansion of the velocity field into wavevector components.

In order that the system be finite, the expansion can be truncated at a wavenumber $k_{\rm max}$ sufficiently large compared to typical dissipation wavenumbers that excitation above $k_{\rm max}$ can be neglected. The real variables $y_n$ at each $\bk$ are the amplitudes of the real and imaginary parts of the flow components perpendicular to $\bk$. The kinetic energy is given by \rf;7; if the fluid density is set to unity. Orthogonal transformations yield other representations in which $E_i$ is a sum of squares. One class of such bases is wavelets, which may be appropriate to describe the intermittency of turbulent flow.

High Reynolds number turbulence maintained in statistically steady state by forcing at large spatial scales and viscous dissipation at much smaller spatial scales exhibits a cascade of energy through intermediate scales, the inertial range. Energy cascade is the most fundamental dynamical feature of the turbulence. It is not a phenomenon of absolute equilibrium, whether Boltzmann-Gibbs or Tsallis. Energy cascade requires nonzero triple correlations of the form $\avv{y_ny_my_r}$, where $n,m,r$ label modes with different wavenumbers or other scale labels. Such correlations are not produced by any flavor of equilibrium statistics based on the kinetic energy: the kinetic energy is a sum of squares and is invariant under $y_n\to-y_n$. The characteristic times of the flow modes associated with energy cascade and other dynamical features decrease with scale size but are finite at all scales.

In any representation where \rf;7; holds, the results of Sec. 2 imply that neither Boltzmann-Gibbs nor $q \ne 1$ Tsallis statistics describes high Reynolds number turbulence. Both statistics yield equipartition of kinetic energy among the degrees of freedom because of the symmetry of \rf;7; in the $y_n$ on each energy shell. An application of Tsallis statistics to turbulence, viable in the sense of fitting certain modal PDFs, must break the symmetry over the degrees of freedom. In some work, individualized values of $q$ and $\be$ are invoked for subsets of the degrees of freedom and/or the kinetic energy is replaced by a modified energy function that is different for different degrees of freedom.

Different $q$ and $\be$ values for two subsets of the flow degrees of freedom may possibly be justified as an approximation if the subsets have dynamical coupling sufficiently weak compared to the strength of their internal dynamics. If the subsets are ranges of wavevector amplitudes in a homogeneous flow, both approximate theory and DNS suggest that this requires separation of the ranges by more than a decade in $k$. If some weakness of coupling criterion is not satisfied, there is no justification for assigning different $q$ and $\be$ values to different degrees of freedom.

The turbulence PDFs that have been approximated in applications of generalized equilibrium statistics have shapes that are easy to fit in a number of ways: the PDFs are smooth and nearly symmetrical with flaring skirts. Justification of a fit from turbulence dynamics is essential for the fitting process is to be considered meaningful.

The applications by Arimitsu and Arimitsu use the link between Tsallis statistics and multifractals \cite{2,3} to determine a single $q<1$ value for the entire system, related to the intermittency behavior associated with a multifractal turbulence model. The multifractal model describes intermittency that increases with decrease of scale size. Multifractal models of turbulence have not been derived from the NS equation but they are supported by theoretical arguments and their parameters can be tuned to agree well with a variety of experimental measurements.

Many of the applications deal with velocity increments across spatial separations. Velocity increments do not constitute a representation basis for a flow. In contrast to the invertible transformation between, say, wavevector components and wavelet amplitudes, the transformation from the velocity field in $x$ space to sums and differences of velocities at two space points is overcomplete. If the velocity sums are struck out, leaving only the increments, the transformation is not invertible. Analysis that derives results by working with velocity increments rests on the implication of corroboration in a representation basis from which the velocity increments are constructed.

In Sections to follow, we shall examine several applications of Tsallis statistics to turbulence. Before that, we shall summarize the behaviors of three systems with a single dynamical variable for which Tsallis statistics is exactly realized: anomalous diffusion in an optical lattice \cite{4}, a Langevin equation with fluctuating parameters \cite{5} and sensitive nonlinear maps \cite{2,3}.

\section*{4. Tsallis Statistics for Single Variables}

Lutz \cite{4} finds that the diffusion of momentum $p$ in an optical lattice treated semiclassically satisfies the Langevin equation
\opn
dp/dt = K(p) + dD(p)/dp + \sqrt{2D(p)}w(t)
\qno;10;
where $K(p)$, $D(p)$, and $dD(p)/dp$ are nonlinear non-stochastic functions of $p$ and $w(t)$ is Gaussian noise with $\avv{w(t)w(t')} = \de(t-t')$. The exact PDF in steady state that follows from \rf;10; is
\opn
P_p(p) = Z^{-1}[1 + \be(q-1)p^2]^{-1/(q-1)}.
\qno;11;
Here $q>1$ and $\be$ are found from $D(p)$ and $K(p)$; $p^2$ comes from the specific forms of $K(p)$ and $D(p)$; $Z$ is the normalization factor.

Lutz notes that the steady state described by \rf;10; and \rf;11; is not an equilibrium state and $\be$ is not a well-defined inverse temperature.

Beck \cite{5} has demonstrated that Tsallis statistics with $q > 1$ are realized by a dynamical system consisting of a degree of freedom that obeys a Langevin equation with fluctuating parameters. Let $y$ satisfy
\opn
dy/dt + \ga y = \si w(t),
\qno;12;
where $\ga$ is a damping constant, $\si$ is a strength parameter, and $w(t)$ is unit white-in-time Gaussian noise. The steady-state PDF of $y$ is Gaussian with variance $1/\be$, where $\be = \ga/\si^2$.

Now suppose that $\ga$ and/or $\si$ fluctuate so that $\be = \ga/\si^2$ has a $\ch^2$ distribution of degree $n$. This means that the PDF of $\be$ is $\prop \be^{n/2-1}\exp(-n\be/2\be_0)$, where $\be_0$ is the mean value of $\be$. Assume that $\be$ fluctuates on time scales infinitely large compared to the $O(1/\ga)$ time required for $y$ to reach equilibrium. Beck shows that the PDF of $y$ has the form
\opn
P(y) \prop 1/[1 + \be'(q-1)y^2]^{-1/(q-1)},
\qno;13;
where
\opn
q = 1 + 2/(n+1) > 1, \quad \be' = 2\be_0/(3-q).
\qno;14;
In comparison with the formulas of Sec. 2, $y^2$ plays the role of an energy.

Although \rf;10; and \rf;12; are dissipative equations, the PDFs \rf;11; and \rf;13; express Tsallis non-dissipative equilibrium if the energy constraint is constructed with the required effective energies.
                                                       
Certain classes of nonlinear one-dimensional maps have a kind of sensitivity to initial conditions, at the onset of chaos, that can be described by Tsallis statistics. The growth of small perturbations in initial conditions is not exponential but instead follows a power law that corresponds to a Tsallis distribution with $1 > q > 0$. Lyra and Tsallis \cite{2} find that the critical attractor for several kinds of maps, including generalized logistic maps and the circle map,  exhibits multifractal behavior in which the singularity strength $\al$ has maximum and minimum values that are related to $q$ in a Tsallis distribution by
\opn
1/(1-q) = 1/\al_{\rm min} - 1/\al_{\rm max}.
\qno;15;

\section*{5. Acceleration of Fluid Elements}

The statistics of pressure forces, viscous forces, and the resulting acceleration of fluid elements in incompressible NS flow has drawn continuing attention for more than 50 years \cite{6,7,8,9}. The behavior of fluid-element acceleration (acceleration along Lagrangian trajectories) can be obtained completely if the distribution of the velocity field is taken to be Gaussian with a specified spectrum. Suppose that the Gaussian velocity field is isotropic, with a wavenumber spectrum that includes a long $k^{-5/3}$ range followed by a rapidly falling dissipation range. Then the acceleration $a$ is dominated by pressure forces, which make a contribution
\opn
\avv{a^2} = c_p\ep^{3/2}\nu^{-1/2}
\qno;16;
to the mean-square $\avv{a^2}$, where $c_p$ is a number nominally $O(1)$, $\ep$ is the average dissipation per unit mass, and $\nu$ is the kinematic viscosity. The viscous forces make a contribution of the same form and nominal magnitude, but the corresponding coefficient $c_v$ is smaller than $c_p$ by a large enough factor that $a$ is strongly dominated by the pressure-gradient contributions.

In realizations of approximately isotropic NS turbulence at large Taylor microscale Reynolds number $R_\la$ by DNS or experiment, intermittency strongly affects the distribution of pressure-gradient forces and causes a small deviation from the $-5/3$ energy spectrum exponent for the inertial range. The effect on the pressure-gradient contribution to $\avv{a^2}$ probably is to make $c_p$ weakly dependent on the ratio $k_d/k_0$ of the dissipation and macro-scale wavenumbers. It is to be expected that the dominance of pressure over viscous contributions to $a$ is enhanced compared to the Gaussian case, especially for contributions to the tails of the PDF of $a$.

Beck \cite{10} constructs Tsallis statistics for the acceleration, with the conclusion that the normalized PDF of $x = a/\avv{a^2}^{1/2}$ is
\opn
P_x(x) = {2\over\pi}{1\over(1 + x^2)^2}.
\qno;17;
The analysis uses a single variable $u$ defined as the difference in the velocity measured at two times along a fluid-element trajectory. The time difference $\ta$ is taken to be the order of the characteristic dissipation time $\ta_\et = (\nu/\ep)^{1/2}$ of the 1941 Kolmogorov theory (K41). The acceleration is measured as $a = u/\ta$.

Beck models the dynamics of $u$ by that of $y$ in the Langevin equation with Gaussian and $\ch^2$ fluctuating parameters whose analysis is summarized in Sec. 4. He states that an excellent fit to experimental data on the PDF of $a$ \cite{11} and on $\avv{a^2}$ \cite{12} is obtained by taking $q=3/2$ and using, instead of $\be'$ in \rf;13;, an $O(\ep^{-1/2}\nu^{-1/2})$ parameter whose precise value is tuned. The choice $q = 3/2$ gives \rf;17;.

In more recent work (private communication and \cite{10}), a fit to normalized pressure gradient component data $x_i=p_i/\avv{p_i^2}$ from approximately isotropic DNS at $R_\la = 381$ \cite{13} is made using log-normal rather than $\ch^2$ statistics in \rf;12;. There is one fitting parameter. The bold simplicity of \rf;17; is lost, but the fit is extended over the range to $P_x(x) \sim 10^{-12}$. The $\ch^2$ and log-normal fits are plotted with the DNS PDFs in Figs. 1 and 2. Fig. 2 shows that neither fit is accurate in the central peak region, which contains most of the weight of $P_x(x)$.

A fluctuating equilibrium environment, of which the log-normal and $\ch^2$ statistics are particular examples, is inconsistent with the turbulent energy cascade. See Sec. 8 for further discussion. The white-noise forcing in \rf;12; conflicts with the fact that turbulence modes of given scale are principally driven by modes of larger scale that have finite and longer characteristic times.

The DNS pressure gradient curves agree well with experimental acceleration data \cite{11} measured in the approximately isotropic central region of flow between counter-rotating disks at reported $R_\la$ values of 200, 690, and 970. The experimental data extend down to about $10^{-6}$ on the vertical axis.

\end{multicols}

\begin{figure}
\hskip 12mm
\epsffile{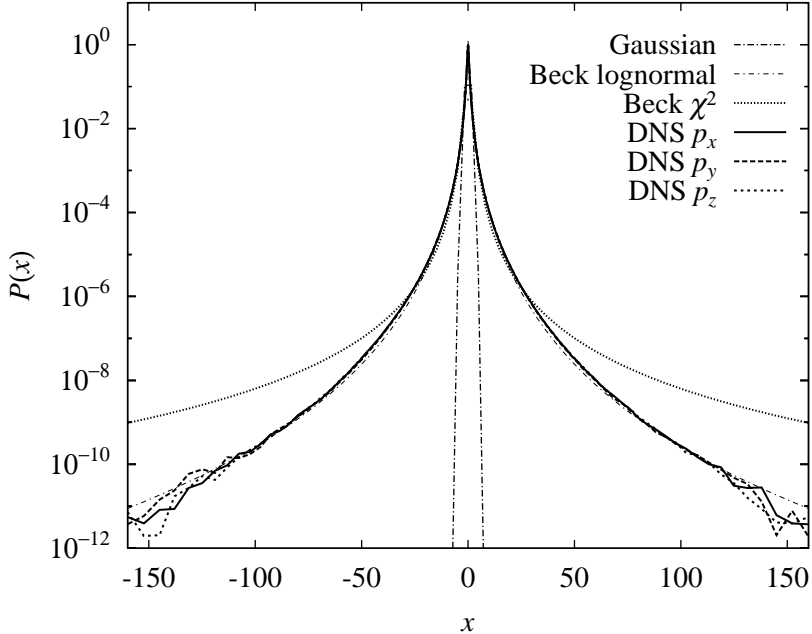}
\hskip -12mm
\caption{
Lin-log plot of normalized PDFs of components of DNS pressure gradient at $R_\la = 381$ together with Beck's $\ch^2$ and lognormal fits. The tails of the DNS PDF curves fit a stretched exponential of form $\exp(-|x|^{0.41})$.}
\end{figure}

\resume

Arimitsu and Arimitsu \cite{3} derive an acceleration PDF from a multifractal cascade model. The starting point is \rf;15;, which they generalize to the form $1/(1-q) = 1/\al_- - 1/\al_+$, where negative $f(\al)$ is admitted and $\al_-$ ($\al_+$) is the smaller (larger) zero of the multifractal spectrum $f(\al)$. The spectrum $f(\al)$ is linked to that of a cascade model for isotropic turbulence. Specifically,
\opn
f(\al) = 1 + (1-q)^{-1}\log_2[1 - (\al -\al_0)^2/(\De\al)^2],
\qno;18;
where $(\De\al)^2 = 2X/[(1-q)\ln2]$ and $q$, $X$, $\al_0$ are determined from the intermittency exponent $\mu$. The multifractal cascade is assumed to proceed in decrement steps of $1/2$ in ``eddy size''. After $m$ statistically independent steps the probabilility $P^{(m)}(\al)d\al$ to find an eddy with singularity in the range $\al$ to $\al+d\al$ is determined as

The acceleration statistics are obtained from the scaling $\de p_m/\de p_0 = (\ell_m/\ell_0)^{2\al/3}$. Here $\de p_m$ is the pressure difference across the distance $\ell_m = 2^{-m}\ell_0$ and $\ell_0$ is a turbulence macroscale. The cascade is terminated at step $n$ chosen so that $\de p_n/\ell_n$ is expected to approximate the pressure gradient.

An additional element in the analysis is ``thermal fluctuations'' that have a Gaussian or $q>1$ Tsallis distribution and are expected to dominate the acceleration PDF for small amplitudes. The total PDF is written as the sum of a contribution from the cascade up to step $n$ and the ``thermal fluctuations'' part. We do not find a physical basis for the ``thermal fluctuations''. If such fluctuations existed, we would expect them to pervade space so that they add everywhere to the field produced by the multifractal cascade. In that case, the PDF of the total field would not be the sum of PDFs from the two contributions to the field.

\end{multicols}

\begin{figure}
\hskip 15mm
\epsffile{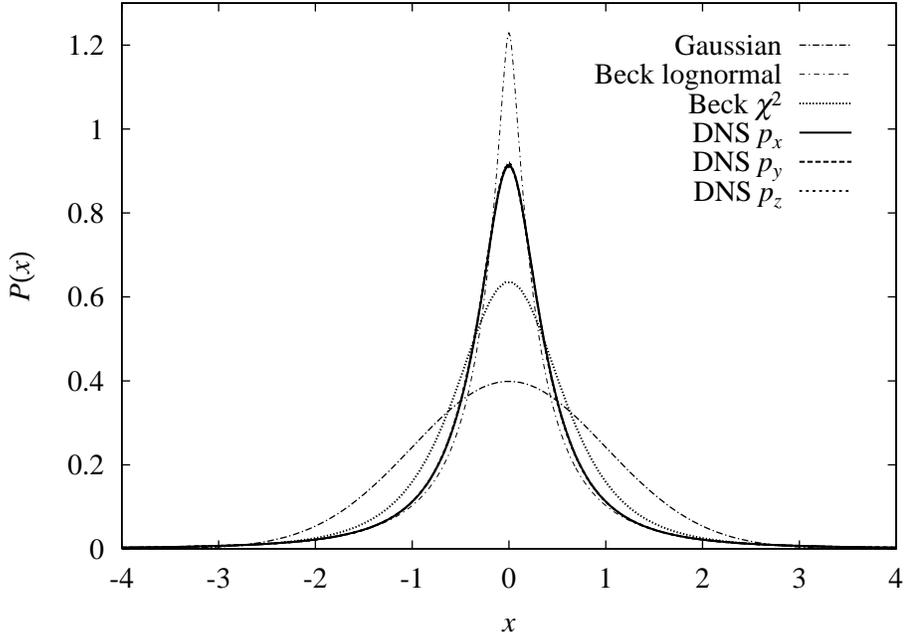}
\hskip -15mm
\caption{
Same as Fig. 1 but only the central regions of the PDFs are shown on a lin-lin plot.}
\end{figure}

\begin{figure}
\hskip 12mm
\epsffile{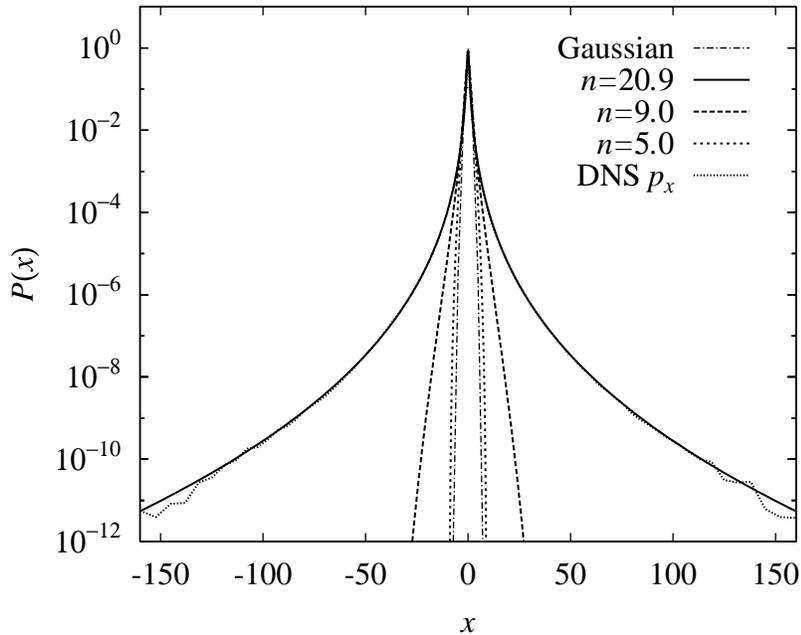}
\hskip -12mm
\caption{
Lin-log plot of PDF of normalized DNS pressure gradient component $p_x/\avv{p_x^2}^{1/2}$ at $R_\la = 381$ compared with the PDF proposed for fluid-element acceleration in [3].      Curves for cascade termination step numbers of $n =$ 20.9, 9.0, and 5.0 are shown. For these fits, $\mu = 0.210$ and $q = 0.3183$.}
\end{figure}

\resume

\opn
P^{(m)}(\al) = [P^{(1)}]^m \prop [1 - (\al-\al_0)^2/(\De\al)^2]^{m/(1-q)}.
\qno;19;

We have calculated acceleration PDFs from the combined multifractal and Gaussian theory as presented in \cite{3}. The results are compared with DNS \cite{13} of pressure gradient in Figs. 3 and 4 for $n = 20.9$, $n= 9$, and $n = 5$. The curves for these $n$ values are obtained with $\mu = 0.210$, $q = 0.3183$. $R_\la = 381$ in the DNS.

The DNS is cyclic in each axis direction and the number of points in each direction is $2^{10} = 1024$. The largest possible value of $\ell_0$ pointing along an axis therefore is $2^9=512$ grid point spacings (this is a distance $\pi$). Thus the upper limit for $n$ is 9. In fact, this limit should be reduced because the forcing extends from the lowest wavenumber $k=1$ to $k=\sqrt{6}$. The multifractal cascade consequently should be taken to start at no more than $512/\sqrt{6} \approx 209$ grid point spacings.

\end{multicols}

\begin{figure}
\hskip 15mm
\epsffile{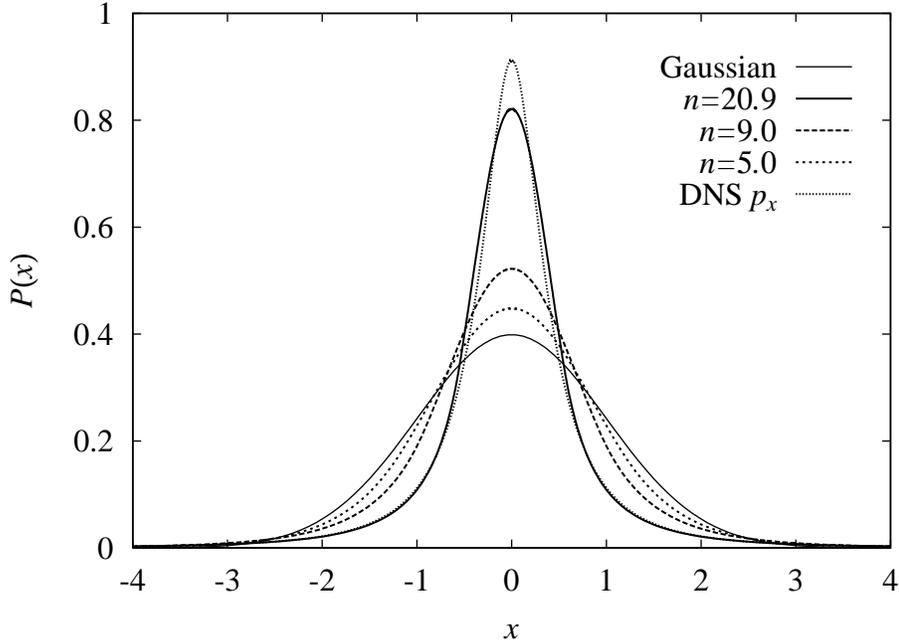}
\hskip -15mm
\caption{
Same as Fig. 3 but only the central regions of the PDFs are shown on a lin-lin plot.}
\end{figure}

\resume

There is substantial scale nonlocality in the turbulence coupling so that there must be a non-asymptotic region at the start of the assumed multifractal scaling range. If we assume that this region starts at approximately 209 grid point spacings and extends for approximately 3 of the nominal cascade halving steps, we get 5 as a plausible value of $n$. If $R_\la$ were increased to 1000, the inertial range would be extended by approximately two scale-size halvings.

Although the value $n=20.9$ is not meaningful for the DNS, we show it because it gives optimum fitting of the theoretical curve to the skirts of the DNS curve. We find that the optimum fit obtained by tuning both $\mu$ and $n$ is insensitive to variation of $\mu$ over the range 0.210 to 0.260 and a corresponding variation of $n$ from 20.9 to 17.0. Over this range the resulting theoretical curve is almost invariant. There appears to be no way to bring the tails of the theoretical curve for $n = 9.0$ or $n = 5.0$ close to the DNS by tuning $\mu$.

The best fit stated in \cite{3} uses $\mu=0.241$, $n=18.2$. This is almost identical with one of the fits we found ($\mu=0.240$, $n=18.0$) in the parameter domain where the result is nearly invariant to fitting changes. Note that $n=18$ would correspond to 1/512 grid point spacing if $\ell_0$ is given the maximum possible value of 512 grid point spacings.

The crucial parameter in the fits is the number of statistically independent cascade steps, which is the maximum value $n$ of $m$ in \rf;19;. In the preceding discussion, we have started with a maximum meaningful value of $\ell_0$ and assumed that the scale decrement in each cascade step is 1/2. Then $n=18$ yields the meaningless minimum scale size 1/512 grid point spacing. Arimitsu \& Arimitsu discuss two alternatives to force the minimum scale size to be the order of one grid point spacing with $n=18$. One is to make $\ell_0$ much larger than the largest meaningful value of 512 grid point spacings. The other is to make the inverse of the scale decrement factor less than 2. It must be stressed that a factor of 2 is already is smaller than the scale ratio needed for approximate dynamical independence of flow modes.

\section*{6. Scale Dependent Tsallis Parameters}

K41 implies that velocity statistics associated with the succession of scales in the inertial range are similar. It has long been known from experiment and suggested by theoretical arguments that this is not the case. Instead, there is intermittency that increases with decrease of spatial scale. One manifestation of the intermittency is the behavior of the velocity structure functions, defined as moments of order $n$ of velocity differences across a distance $r$. As $r$ takes successively smaller values in the inertial range, the structure functions show an increasingly rapid growth with $n$, in conflict with K41 similarity. The growth of the structure functions expresses anomalous scaling exponents. A number of theoretical models have been explored, but almost nothing has actually been deduced from the NS equation.

Beck has studied a series of models of the longitudinal velocity difference $u$ (difference in the velocity component pointing along $\br$). We shall consider them in more or less chronological order. In the first of these models \cite{14}, the dynamics of $u$ is related to a Langevin equation with non-stochastic damping and a stochastic forcing term:
\opn
du/dt + \ga u = f(t).
\qno;20;
Here $f(t)$ is what Beck calls deterministic chaotic noise with typical time scale $\ta \ll 1/\ga$. The statistics of $f$ is such that the PDF of $u$ acquires a skewness. The actual qualitative physics of transfer into modes of given scale is that the transfer comes mostly from modes of larger scale with longer dynamical times. This conflicts with $\ta \ll 1/\ga$ in the model equation \rf;20;.

The result is that for $c/\sqrt{R_\la}\ll 1/\hu$ the PDF of $\hu = u/\sqrt{\avv{u^2}}$ has the form
\opn
p(\hu) = {1\over Z_q}\left[ 1 + \be\tq\left( {\textstyle\frac{1}{2}} \hu^2 - {c\over\sqrt{R_\la}} \left( \hu - {\textstyle\frac{1}{3}} \hu^3 \right)\right)\right]^{-1/\tq},
\qno;21;
where $q$ is a Tsallis entropy parameter, $\tq=q-1$, $R_\la$ is the Taylor microscale Reynolds number of the turbulence, and $c$ is a positive $O(1)$ constant. Higher-order terms in $c$ prevent a change of sign of $p(\hu)$ at large negative $\hu$ and make $p(\hu)$ normalizable. The value of $\be$ is adjusted to give $\hu$ variance 1, which requires $\be = 2/(5-3q)$ for large $R_\la$.

How is \rf;21; obtained from \rf;20; and how is $q$ determined? There are a number of steps. The small parameter $\ga\ta$ is identified with $1/R_\la$ by the following argument: since $\ga$ is a damping, it should be proportional to the underlying viscous damping coefficient $\nu$; correct dimensionality then requires a length scale, which is set equal to the Taylor microscale $\la$, so that $\ga = \nu/\la^2$. Also it is assumed reasonable to take $\ta = \la/\sqrt{\avv{u^2}}$. Thus $\sqrt{\ga\ta} = c/\sqrt{R_\la}$ with $c = O(1)$.

The parameter $q-1$ is set according to the number of cascade steps between $r$ and the K41 dissipation length $\et = (\nu^3/\ep)^{1/4}$. A cascade step is taken to be a factor of 2 in $1/r$. This leads to
\opn
1/(q-1) = O(1) + \log_2(r/\et).
\qno;22;
Finally $O(1)$ is set to 1 because that is said to give agreement with experimental data.

Beck uses \rf;21; to calculate moments of the normalized velocity difference $\hu$ (normalized structure functions) and from them calculates scaling exponents for the normalized structure functions.

Some comments on \rf;20;--\rf;22;: In K41, the characteristic time at scale $r$ in the inertial range is $O(r/v_r)$ where $v_r = (r/\et)^{1/3}v_\et$ and $v_\et = (\ep\nu)^{1/4}$. The eddy damping rate, which measures the rate of cascade to smaller $r$, is $O(v_r/r)$. Scales in the inertial range see only the eddy damping and do not see $\nu$. A physical cascade step must be more than a factor of 2 in $1/r$; there is still substantial energy transfer at a factor of 4. The maximum number of effective cascade steps was discussed in Sec. 5 and will be discussed further in Sec. 7.

The skewness $\avv{\hu^3}/\avv{\hu^2}^{3/2}$ is independent of $r$ in the inertial range of K41. It measures the rate of energy cascade. There is no dependence on $R_\la$. Experiment and DNS indicate that the skewness is very nearly independent of $r$ and $R_\la$ in actual intermittent turbulence. The same conclusion comes from several models of intermittency, including the beta model and multifractal models \cite{15}.

We noted in Sec. 3 that the energy cascade cannot be captured if the kinetic energy, which is a sum of squares in $\bk$ or wavelet bases, is used in the Tsallis formalism. In \rf;21;, skewness is given to $\hu$ by adding linear and trilinear terms to the actual kinetic energy. This procedure requires examination.

Suppose that $\ga$ in \rf;12; is non-stochastic but $\si$ is stochastic and constant in time. Then the PDF of $y$ is obtained by a weighted integration over the variance parameter of a Gaussian PDF. Very wide classes of PDF can be generated in this way. See Sec. 8 for further discussion. A Tsallis PDF with $q>1$ is generated by $\ch^2$ statistics for $1/\si^2$, as described in Sec. 4. No choice of weight function can give $y$ a skewness. To obtain skewness, the weighted white noise $\si w(t)$ must be replaced by a skewed forcing with finite correlation time. This is taken as the $f(t)$ of \rf;20;.

We have difficulty accepting the relation between turbulence and these manipulations with single-degree-of-freedom models. Problems with characteristic times have already been noted. The inclusion of skewness in the models seems to be motivated by the empirical fact that the PDF of $\hu$ is skewed, rather than any dynamical link to turbulence structure.

In later work \cite{16}, \rf;21; is modified and the dependence on $R_\la$ is dropped; $c/\sqrt{R_\la}$ is replaced by $c\sqrt{\ga\ta}$ and it is stated that a match to experimentally measured skewness requires $\sqrt{\ga\ta}$ to be about 0.1. It is stated that $c$ may vary with the experiment and may depend on $q$. Also, it is implied that the variation of $q$ with $r$ needs to be determined {\it a posteriori}. An additional parameter $\al$ is introduced yielding the effective energy
\opn
E(u) = {\textstyle{1\over2}}|u|^{2\al} - c\sqrt{\ga\ta}\,\,\sgn(u)(|u|^\al - {\textstyle{1\over3}}|u|^{3\al}).
\qno;23;
It is stated that $\al < 1$ corresponds to a fractal phase space, in analogy to the beta-model picture of active eddies confined to a fraction of space.

The complicated theoretical apparatus thus constructed can give matches to the normalized PDFs of $u$ measured in the central region of a Taylor-Couette flow at macroscale Reynolds number $R = 540\,000$. The flow is between a rotating inner cylinder and a stationary outer cylinder. The best fits are obtained by the empirical procedure of taking $\al = 2-q$, $c\sqrt{\ga\ta} = 0.124(q-1)$, and then choosing $q$ to give minimum mean-square deviation from the measured PDF. Once these expressions for $\al$ and $c\sqrt{\ga\ta}$ are fixed, the only free parameter is $q$. If the parameter $\al$ is not invoked, the range of $1/(q-1)$ needed for best fits to experimental data increases and there is some loss of accuracy in the best fits. The range of $q-1$ used in fitting the measured curves is from 0.055 at $r/\et = 14\,400$ to 0.168 at $r/\et = 11.6$, corresponding to a range of $1/(q-1)$ from 18.2 to 5.95. If the parameter $\al$ is not employed, the range increases.

As already noted, the velocity difference $u$ does not belong to a valid representation basis for the flow. Analysis with given $q$ and $\be$ must apply to appropriate modes in a valid underlying basis. The implication is that the given $q$ and $\be$ must describe all statistics dominated by those modes, including but not limited to the statistics of transverse velocity differences.

The modeling that yields \rf;23; has been refined \cite{17} by appeal to extended self-similarity (ESS) to determine $\be$ and thereby reduce the number of free parameters. Close matches to experimental estimates of scaling exponents are obtained. An effort in a different direction \cite{18} finds that the Tsallis entropies of subsystems can be made quasi-additive by adjustment of the $q$ values for the subsystems. This leads to another formula for the variation of $q$ with $r$.

We have two  principal difficulties with models in which $q$ varies with $r$. The first is that the analysis appeals to fluctuations of effective equilibrium temperatures. Equilibrium temperatures, fluctuating or not, are inconsistent with the dominance, in the turbulence dynamics, of the strong and on-average unidirectional multistep energy cascade. The second difficulty, noted in Sec. 3, is that modes whose scales differ by less than a decade have substantial dynamical coupling so that assigning them different $q$ values seems unjustified.

\section*{7. Multifractal Analysis of Velocity Fluctuations}

Arimitsu and Arimitsu \cite{3} use the relation of Tsallis statistics to multifractal processes to develop a multifractal model of high Reynolds number isotropic turbulence that gives detailed predictions of scale-dependent statistics. In addition, scaling exponents of velocity differences and other properties are derived and compared with the approximately isotropic DNS of \cite{13}. Other, empirically developed and tuned, multifractal models have previously been found to agree well with a variety of measurements \cite{19}. 

As noted in Sec. 5, the starting point of the analysis by Arimitsu and Arimitsu is the relation \rf;15; between $q$ and the extremes of the multifractal spectrum $f(\al)$ of a one-dimensional nonlinear map that displays sensitivity to initial conditions. This relation is generalized and then applied to the multifractal spectrum of a turbulence model. An explicit value for $q$ is found in terms of the intermittency exponent $\mu$ of the model.

Some remarks on the physical nature of the inertial-range cascade: Viscosity is not felt directly at a scale $r$ well within the inertial range. Instead the cascade of kinetic energy is characterised by a transfer rate $\mit{\Pi(r)}$. $\mit{\Pi(r)}$ is not positive-definite, and the DNS shows that it has a finite probability of negative values \cite{13}. In contrast, all spatial averages of viscous dissipation are positive. This contrast is recognized in \cite{19}. Finite time-scales are associated with the cascade process at the scale $r$; there is not an instantaneous relation between cascade rate over a domain of size $r$ and dissipation averaged over the domain. Moreover the energy cascade is sufficiently diffuse that there is significant transfer at scale ratios 4 and larger.

Arimitsu and Arimitsu use the multifractal formalism described in Sec. 5 to fit scaling exponents of velocity differences presented in \cite{13}. Both longitudinal and transverse differences are fitted. They find that an optimum fit over a wide range of space intervals $r$ is obtained by taking $\mu = 0.327$ and $n$ values that range from 21.5 to 4.00 for $r/\et$ values that range from 2.38 to 1220.

Maximum admissible $n$ values for various $r/\et$ values  follow from the discussion in Sec. 5. We noted there that the effective number of grid point spacings at the beginning of the unforced cascade range is $\approx 209$; that there is substantial scale nonlocality in the turbulence coupling; and that there must be a non-asymptotic region at the beginning of the scaling range. Since $\et$ in the DNS is approximately 1 grid point spacing, this leads to the conclusion that the maximum admissible $n$ value for $r \approx \et$ is less than 9. Extension of this argument to other $r$ values gives $n =$ 8, 7, 6, 5, 4, 3, 2,  1, 0 as upper limits for $r/\et =$ 1.95, 3.91, 7.81, 15.6, 31.3, 62.5, 125, 250, 500.

All of the $n$ values used in the optimum fitting in \cite{3} greatly exceed the maximum admissible $n$ values given above. We are thereby stymied in considering further the fits to velocity-difference statistics and scaling exponents of structure functions. We cannot assess the proposal that there are two multifractal scaling ranges separated by a crossover value of $r$.

\section*{8. Concluding Remarks}

The inviscid conservation of kinetic energy in incompressible 3D NS flow gives important information about the energy cascade process at large Reynolds number and the associated skewness of the velocity field. Energy conservation fails to provide information about higher statistics -- in particular, the development of intermittency in the cascade. A number of mechanisms that affect higher statistics are known. Vortex stretching implied by the Kelvin circulation theory and the existence of several kinds of instability play important roles. However, it is not known how to put all these things together into a deductive theory of turbulence statistics. Instead, there are a variety of models of higher statistics that have meager or nonexistent deductive support from the NS equations but can be made to give good fits to experimental measurements.

A straightforward application of Tsallis statistics to turbulence suffers the same basic limitation as Boltzmann-Gibbs statistics: the equilibria are based solely on the kinetic energy of the flow. It was noted in Sec. 3 that all the degrees of freedom enter symmetrically into the kinetic energy and therefore are equally excited in equilibrium, whether Boltzmann-Gibbs or Tsallis. Two strategies that have been used to break the symmetry are the construction of effective modal energies that differ from the actual kinetic energy and the use of different values of the nonextensivity parameter $q$ for different parts of the system. Some of the work that uses these strategies is described in Secs. 5--7. As explained in Sec. 3, the result is no cascade of energy between groups of modes at different scales, whether or not skewness appears in the PDFs.

An absurd, but perhaps illuminating, manipulation of effective energies that breaks the symmetry over modes can be performed within Boltzmann-Gibbs statistics. Any desired PDF $P_n(y_n)$ can be realized for a mode $y_n$ by taking $-\be e_n(y_n) = \ln P_n(y_n)$ in \rf;6; to determine $e_n(y_n)$. $P_n(y_n)$ can be constructed to include skewness, but again the result does not describe the energy cascade. Energy transfer requires nonzero correlations of form $\avv{y_ny_my_r}$, where $n,m,r$ label modes with differing wavenumbers or other scale labels.

Beck and his coworkers treat velocity differences across distances $r$ as dynamic flow variables, which they are not. A problem with their approach is that each velocity difference is modeled by a modified Langevin equation with white-noise forcing. This conflicts with the fact that the driving of modes of given spatial scale is dominated by modes of larger scale that have finite and longer characteristic times.

Arimitsu \& Arimitsu generalize single-variable multifractal attractors for which Tsallis statistics is exact and propose a link to multifractal turbulence models. A key quantity in the fits to DNS statistics is the number of nominal ``eddy size'' halvings. Impressive fits are made if this number is freely disposable, but they are destroyed if the number is limited to meaningful values.

The PDFs that have been fitted in applications of Tsallis statistics to NS turbulence have benign shapes: smooth and nearly symmetrical with flaring skirts. Such shapes can be generated in many ways. For example, consider the fits to acceleration PDF examined in Sec. 5. With all questions of consistency and validity put aside, the fact is that both Beck and Arimitsu \& Arimitsu present fits to the entire range of the DNS PDF. Yet the fitting algorithms and physical models used by these authors are very different. Fits to PDFs of velocity increments, and scaling exponents, have been obtained from a variety of models constructed by Beck. More challenging would be shapes like the strongly asymmetrical and sharply cusped PDF of velocity gradient in Burgers turbulence forced at low wavenumbers \cite{20}.

Beck and Cohen \cite{21} consider general ensembles of Boltzmann-Maxwell distributions over which $\be$ is itself stochastically distributed. The $\be$ distributions explicitly examined include $\ch^2$ and log-normal. If the energy is a sum of squares of modal amplitudes, the PDFs of mode amplitudes over such ensembles can be given protean shapes, with a smooth central peak and flaring skirts, by adjusting the $\be$ distribution. A wide variety of empirical PDFs can be fitted in this way. In order to make an application to cascading turbulence, the interaction of a mode like, say, a $\bk$ or wavelet amplitude with the rest of the system would have to be modeled as an interaction with a heat bath whose $\be$ fluctuates. An ensemble of equilibria with heat baths is inconsistent with the energy cascade, in which there is a strong, on-average unidirectional, flow of energy from larger to smaller scales. Strong here means that the cumulative mean energy flow through, say, a double octave of wavenumbers in one local characteristic time is the order of the mean energy in the double octave.

Multifractal cascade models raise the general issue of distinction between what is descriptive of physical behavior and what can be used for analysis of data. In \cite{19} experimental data are analyzed to yield generalized dimensions $D_p$ and the corresponding multifractal spectrum over the range $-20 \le p \le 20$. Then the multifractal model is tuned to give an optimum (and excellent) fit to the data. Multifractal models may or may not express well the cascade physics at large but finite Reynolds numbers.

\section*{Acknowledgerments}

It is a pleasure to acknowledge helpful conversations and correspondence with T.~Arimitsu, C.~Beck, E.~Bodenschatz, J.~L.~Lebowitz, and H.~L.~Swinney.
\vskip 4mm
T.~G.'s work was supported by a Grant-in-Aid for Scientific Research (C-2 12640118) from the Japan Society for the Promotion of Sciences. R.~H.~K.'s work was supported by the Center for Nonlinear Studies, Los Alamos National Laboratory, U.~S.~A., under subcontract B2933-001-96.

\opbib

\vskip -5mm

\bb{1} R\'{e}nyi, A. {\it Proc. 4th Berkeley Symp. Maths. Stat. Prob.} (1961) 479; Tsallis, J. Stat. Phys. 52 (1988) 479.

\bb{2} M. L. Lyra and C. Tsallis, Phys. Rev. Lett. 80 (1998) 53.

\bb{3} T. Arimitsu and N. Arimitsu, arXiv:cond-mat/0203240; T. Arimitsu and N. Arimitsu, J. Phys.: Condensed Matter 14 (2002) 2237; T. Arimitsu and N. Arimitsu, arXiv:cond-mat/0210274.

\bb{4} E. Lutz, arXiv:cond-mat/0210022.

\bb{5} C. Beck, Phys. Rev. Lett. 87 (2001) 180601.

\bb{6} W. Heisenberg, Z. Physik 124 (1948) 628.

\bb{7} G. K. Batchelor, Proc. Cambr. Phil. Soc. 47 (1952) 359.

\bb{8} A. S. Monin and A. M. Yaglom, {\it Statistical fluid mechanics} (M.I.T. Press, Cambridge, Mass., 1975), Sec. 18.3.

\bb{9} A. Tsinober, P. Vedala and P. K. Yeung, Phys. Fluids (2001) 1974. This paper cites many other studies.

\bb{10} C. Beck. Phys. Lett. A 287 (2001) 240; arXiv:cond-mat/0212566.

\bb{11} G. A. Voth, K. Satyanarayan, and E. Bodenschatz,  Phys. Fluids 10 (1998) 2268; A. La Porta, G. A. Voth, A. M. Crawford, J. Alexander, and E. Bodenschatz, Nature 409 (2001) 1017; G.~A.~Voth, A.~La~Porta, A.~.M.~Crawford, J.~Alexander, and E.~Bodenschatz, J.~Fluid~Mech. 469 (2002) 121; A.~M.~Crawford, N.~Mordant, and E.~Bodenschatz, arXiv:physics/0212080.

\bb{12} P. Vedula and P. K. Yeung, Phys. Fluids 11 (1999) 1208.

\bb{13} T. Gotoh, D. Fukayama and T. Nakano, Phys. Fluids 14 (2002) 1065; T.~Gotoh and T.~Nakano (2002) unpublished.

\bb{14} C. Beck, Physica A 277 (2000) 115.

\bb{15} U. Frisch, ``Turbulence,'' (Cambridge University Press, Cambridge, England, 1995).

\bb{16} C. Beck, G. S. Lewis and H. L. Swinney, Phys. Rev. E 63 (2001) 035303.

\bb{17} C. Beck, Physica A 295 (2001) 195.

\bb{18} C. Beck, Europhys. Lett. 57 (2002) 329.

\bb{19} C. Meneveau and K. R. Sreenivasan, Phys. Rev. Lett. 13 (1987) 1424.

\bb{20} W. E. K. Khanin, A. Mazel and Y. Sinai, Phys. Rev. Lett. 78 (1997) 1904; R.~H.~Kraichnan, Phys. Fluids 11 (1999) 3738.

\bb{21} C.~Beck and E.~G.~D. Cohen, arXiv:cond-mat/0205097.

\clbib

\end{multicols}

\end{document}